# Unconventional critical behavior in quasi-one-dimensional $S = 1$ chain NiTe$_2$O$_5$

Jun Han Lee[1], Marie Kratochvílová[2,3], Huibo Cao[4], Zahra Yamani[5], J. S. Kim[6], Je-Geun Park[2,3], G. R. Stewart[6], Yoon Seok Oh[1,*]

[1]*Department of Physics, Ulsan National Institute of Science and Technology, Ulsan 44919, Republic of Korea*
[2]*Center for Correlated Electron Systems, IBS, Seoul, Republic of Korea*
[3]*Dept. of Phys. & Astro., Seoul National University, Seoul, Republic of Korea*
[4]*Neutron Scattering Division, Oak Ridge National Laboratory, Oak Ridge, TN 37831, USA*
[5]*Canadian Neutron Beam Centre, Chalk River, ON, Canada*
[6]*Dept. of Physics, University of Florida, FL, USA*
*corresponding author: ysoh@unist.ac.kr

**Abstract**

Here we report a new quasi-one-dimensional $S = 1$ chain compound NiTe$_2$O$_5$. From the comprehensive study of the structure and magnetic properties on high quality single crystalline NiTe$_2$O$_5$, it's revealed that NiTe$_2$O$_5$ undergoes a transition into an intriguing long-range antiferromagnetic order at $T_N$ = 30.5 K, in which longitudinal magnetic moments along the chain direction are ferromagnetically ordered, while their transverse components have an alternating ferromagnetic-antiferromagnetic coupling. Even though the temperature dependence of magnetic susceptibility represents an archetypal anisotropic antiferromagnetic order, we found that critical behavior of unconventional nature with $\alpha' \sim 0.25$ and $\beta \sim 0.18$ is accompanied by the temperature evolution of the antiferromagnetic order parameter.

Quantum-dynamic as well as thermodynamic tunability for competing interactions lead to a phase transition, where the physical quantities exhibit critical behavior following the power law [1]. Plenty of exotic phenomena have been observed near the (quantum) critical regime of correlated systems, such as the atmospheric dynamics of water vapor [2], superfluid density of two dimensional superconductor [3], surface critical behavior in topological phase [4], Higgs mode in quantum antiferromagnet [5], quantum criticality in multiferroics [6], quantum phase transition in two-dimensional electron system [7, 8], etc. Despite the variety of complex interactions and correlations, the critical behavior only depends on global properties such as the spatial dimension, the symmetry of the order parameter, and the range of interaction rather than microscopic details [1, 9]. Feature of the critical behavior, which is described by a set of critical exponents and scaling functions in the power law, provides the foundation to categorize the collective phenomena as the universality class and to understand the origin of the ground state as well as the phase transition. For example, depending on the space and spin dimensions, magnetic systems are classified as to the universality class depending on their critical exponent $\beta$ of order parameters as follows: for 2D-Ising, $\beta = 1/8$; 3D-Ising, $\beta = 0.33$; 3D-XY, $\beta = 0.35$; 3D-Heisenberg, $\beta = 0.36$; and mean field, $\beta = 0.5$. However, recently, unconventional critical behavior has been observed in strongly correlated electron systems such as superconducting heavy fermion UIr [10] and U(Rh,Ge)$_2$ [11], weak ferromagnetic BaIrO$_3$ [12], organic ferromagnetic TDAE-C$_{60}$ [13], and itinerant ferromagnetic Sr$_{1-x}$Ca$_x$RuO$_3$ [14].

Here we introduce a new quasi-one-dimensional chain compound NiTe$_2$O$_5$, in which spin-1 of Ni$^{2+}$ ions form a one-dimensional chain structure through NiO$_6$ octahedra's edge-sharing as shown in Figure 1(a). For magnetic one-dimensional chain systems, it is known that a magnetic order parameter, described as one-dimensional order parameter with discrete symmetry, can only manage to order at $T = 0$ while remaining disordered at all finite temperatures. On the other hand, two-dimensional order parameters in the one-dimensional chain cannot be ordered even at $T = 0$, not to mention ordering at $T \neq 0$ [15]. However, often finite interchain exchange coupling prohibits the genuine one-dimensional intrachain coupling and results in a long-range order at $T \neq 0$ [16]. Comparing with the reported (quasi) one-dimensional spin systems Sr$_2$CuO$_3$ (3.549 Å)[17] and Y$_2$BaNiO$_5$ (5.7613 Å)[18], NiTe$_2$O$_5$ has a comparable or larger interchain distance of 5.957 Å between Ni$^{2+}$ ions, and could be considered as a quasi-one-dimensional chain. We find that NiTe$_2$O$_5$ undergoes a long-range antiferromagnetic (AFM) order with an archetypical anisotropic AFM anomaly at $T_N = 30.5$ K, and the AFM order parameter develops with intriguing unconventional critical behavior.

Polycrystalline and single crystalline specimens of NiTe$_2$O$_5$ were prepared using the solid state reaction method and the flux growth method, respectively. For the synthesis of the polycrystalline sample, high purity powders of NiO (99.998%) and TeO$_2$ (99.99%) were weighed with a stoichiometric molar ratio of NiO:TeO$_2$ = 1:2. The weighed powder was mixed and pressed in a cylindrical mold. The pressed pellet was sintered at 685 °C for 12 hours. This powder sintering process was repeated twice with intermediate grindings. Powder X-ray diffraction (XRD) experiment was subsequently performed using a Bruker D8 Advances, and crystallographic structure of the synthesized polycrystalline NiTe$_2$O$_5$ was confirmed by Rietveld refinement as shown in Figure 1(c) [19]. The calculated and experimental diffraction patterns are well fit with a reliability factor of $\chi^2 = 1.68$. The determined crystallographic structure and parameters of NiTe$_2$O$_5$ are summarized in Table I. NiTe$_2$O$_5$ has orthorhombic structure of *Pbnm* (#62) space group with the *a*, *b*, and *c* lattice constant of 8.4418, 8.8663, and 12.1203 Å, respectively. The edge-sharing of distorted NiO$_6$ octahedra forms a one-dimensional chain structure along the *c*-axis with $S = 1$ of Ni$^{2+}$ ion (Figure 1(a)). The NiO$_6$ chains are arranged in a distorted square lattice, and the residual Te$^{4+}$ and O$^{2-}$ ions occupy interspace of the chains (Figure 1(b)).

Figure 2(a) shows the temperature dependence of dc magnetic susceptibility $\chi_{dc}(T)$ of polycrystal and *a*, *b*, *c* crystallographic axis of the single crystal, which presents the archetypal AFM anomaly with distinct magnetic anisotropy with respect to the chain direction. With decreasing temperature, the magnetic susceptibility gradually increases following the Curie-Weiss law and starts to deviate from the mean-field behavior $\chi_{dc}^{-1}(T) \propto T$ below ~ 100 K. As shown in the inset of Figure 2(a), the Curie-Weiss fitting between 200 K and 350 K for $\chi_{dc}^{-1}(T)$ along the *c*-axis determines the Curie-Weiss temperature of $\theta_{CW} = -8.87$ K and estimates the intra-chain exchange coupling constant of $J = 3k_B\theta_{CW}/zS(S+1) = -0.57$ meV, where $z = 2$ is the number of nearest-neighbor ions in the chain for the exchange interaction. The deviation from $\chi_{dc}^{-1}(T) \propto T$ is attributed to short range correlations near the ordering temperature. At $T_N = 30.5$ K, NiTe$_2$O$_5$ undergoes a clear long range AFM order accompanied by the distinct magnetic anisotropy for *a*, *b*, *c*-axis. Along the *c*-axis parallel to the chains, the dc magnetic susceptibility (red line) drastically decreases down to 1 % of $\chi_{dc}(T = T_N)$ at $T = 2$ K. On the other hand, *a*- (green line) and *b*-axis (blue line) magnetic susceptibilities manifest negligible/reduced temperature dependences below $T_N$. The magnetic anisotropy of $\chi_{dc}(T < T_N)$ indicates that the Ni$^{2+}$ spins are nearly aligned along the *c*-axis parallel to the chain below $T_N$. The thermodynamic property was studied by measurements of temperature dependent specific heat $C(T)$ from 1.8 to 300 K (Figure 2(b)). A sharp lambda-shaped anomaly associated with the second-order AFM transition is observed at $T_N$. In order to estimate thermodynamic contributions of the magnetic degree of freedom, phonon contribution to the $C(T)$ was evaluated by using both the experimentally measured $C(T)$ of nonmagnetic $MgTe_2O_5$ (Figure S1 of Supplemental Material [21]) [22], and the Debye model with various Debye temperatures. Both $C(T)$ of nonmagnetic $MgTe_2O_5$ and scaled $C(T)$ with the Lindemann factor [23,24] of $L = 0.9630$ exceed the $C(T)$ of the $NiTe_2O_5$ above $T_N$ (see Supplemental Material [21]). Instead, the Debye approximation employing three Debye temperatures ($T_{D1} = 935$ K, $T_{D2} = 605$ K, $T_{D3} = 224$ K) results in the best fit over a wide temperature range above $T_N$ (light blue curve in Figure 2(b)). By subtracting off the calculated phonon specific heat of the Debye model, temperature dependence of the magnetic specific heat $C_{mag}(T)$ is estimated and $C_{mag}(T)/T$ (black dots in Figure 2(c)) is integrated with the temperature. A red solid line displays the temperature evolution of the magnetic entropy of $NiTe_2O_5$. The $S = 1$ system should have a total magnetic entropy of $R\ln(2S+1) = 9.13$ J/K mol (red dashed line in Figure 2(c)), but the integrated magnetic entropy exceeds 9.13 J/K mol as shown by the red line in Figure 2(c). The excess is attributed to residual lattice contribution in the estimated magnetic specific heat. In any case, we believe the full magnetic moment of $Ni^{2+}$ ions is completely ordered at $T_N$.

The magnetic structure of $NiTe_2O_5$ was determined by powder and single crystal neutron diffraction experiments, which have been performed in the C5 beamline at Canadian Institute for neutron Scattering and in the HB-3A four-circle diffractometer at Oak Ridge National Laboratory (ORNL), respectively. For the single crystal neutron diffraction experiment, a 90 mg high quality $NiTe_2O_5$ single crystal was prepared (see inset of Figure 3(c)). Figure 3(a) and (b) compare the experimental and calculated results of the powder neutron diffraction at 300 and 2 K. At $T = 2$ K, distinctive Bragg peaks are observed at the following angles of $\theta = 15.9$ °, 16.8 °, 36.43 °, 36.45 °, 37.62 °, and 37.63 °, which correspond to the magnetic Bragg peaks of (0 1 0), (1 0 0), (2 0 1), (1 2 0), (0 1 3), and (2 1 0), respectively. These additional peaks originate from the AFM order. Figure 3(c) plots the calculated versus the observed squared structure factors from the single crystal neutron diffraction data collected at 4 K. For the magnetic Rietveld refinement, we considered eight possible magnetic symmetries that were calculated by the magnetic symmetry analysis tools at the Bilbao Crystallographic Server [25]. We found that the *Pbnm* magnetic symmetry best fits the data, where the unweighted single-crystal R-factor is $R_f$ = 5.32 %. According to the refinement, the total magnetic moment of NiTe$_2$O$_5$ is 2.18(3) $\mu_B$ per Ni$^{2+}$ ion, which is in perfect agreement with the expected value for $S$ = 1 with a $g$ factor of 2.2 and the $c$-axis (longitudinal) component of the magnetic moment is $|m_c| = m_\parallel = 2.15(3)\ \mu_B$ as expected from the magnetic anisotropy of $\chi_{dc}(T)$. In contrast to the AFM negative $\theta_{CW}$ of −8.87 K under the Curie-Weiss law, the magnetic Rietveld refinement on the single crystal neutron diffraction reveals that $m_\parallel$ are ferromagnetically aligned along the chain but antiferromagnetically coupled between the chains (see Figure 3(d)). In addition, the Ni$^{2+}$ spins also have transverse component $m_\perp$ by canting away from the chain direction and the $m_\perp$ are arranged with left-right-right-left configuration along the chain as shown in Figure 3(d) and (e), where $m_a = 0.27(5)\ \mu_B$, $m_b = 0.27(5)\ \mu_B$, and $|m_\perp| = (m_a^2 + m_b^2)^{\frac{1}{2}} = 0.38(6)\ \mu_B$. Thus, along the chain, $m_\parallel$ are ferromagnetically ordered, but $m_\perp$ have alternating ferromagnetic and antiferromagnetic nearest-neighbor coupling. Further detailed studies on exact local spin correlation and model Hamiltonian are required.

In the vicinity of the phase transition, physical quantities exhibit critical phenomena, which are represented by power-laws and quantitatively characterized by exponents of the power laws, so-called critical exponents. A set of the critical exponents can clarify the universality class of phase transitions without knowing a microscopic picture of complex interactions and correlations. We have investigated the universality class of the intriguing spin configuration in NiTe$_2$O$_5$ with critical behavior of $C_{mag}(T)$ and temperature dependent development of magnetic Bragg peak (100) intensity $\sqrt{I_{(100)}(T) - I_0}$. As shown in Figure 4(a) and (b), $C_{mag}(T)$ and $\sqrt{I_{(100)} - I_0}$, where $I_0$ = 8.43733, follow power laws behavior below $T_N$. From the slopes in the logarithmic plots for $T_N$ and $T_N \pm 0.1$ K within an error of the measured transition temperature $T_t$, we determined that the critical exponents of $\alpha'$ in $C_{mag}(T) \sim [(T_t - T)/T_t]^{-\alpha'}$ and $\beta$ in $I_{(100)} \sim [(T_t - T)/T_t]^{2\beta}$ are $\alpha'$ = 0.14, 0.25, 0.31 and $\beta$ = 0.17, 0.18, 0.20 for $T_t$ = $T_N$ – 0.1, $T_N$, $T_N$ + 0.1 K, respectively. As shown in Figure 4(c), $\beta$ = 0.18 presents the best fit between the experimental data and the power law fitting result for $I_{(100)} \sim [(T_N - T)/T_N]^{2\beta}$. Notes that the studies of single crystal neutron diffraction and $\chi_{dc}(T)$ seemingly reflect an archetypical three-dimensional AFM order, which can belong to either 3D Ising of $\beta$ = 0.33 or 3D Heisenberg of $\beta$ = 0.36. In contrast, it is discovered that the experimentally estimated critical exponents of $\alpha' \sim 0.25$ and $\beta \sim 0.18$ show a clear discrepancy with the conventional universality class of 3D Ising ($\alpha' = 0.11$ and $\beta = 0.33$) and 3D Heisenberg ($\alpha' = 0.12$ and $\beta = 0.36$). The unconventional critical behavior has been observed in the two-dimensional magnet within a window $0.1 \leq \beta \leq 0.25$ [26], near the tricritical point of La$_{0.6}$Ca$_{0.4}$MnO$_3$ with $\beta = 0.25$ [27], and in a two-dimensional conductor with $\beta = 1$ [28], etc. However, understanding of the unconventional critical behaviors in the one-dimensional system is still unclear, and further experimental and theoretical studies are required. Since the magnetic ground state of NiTe$_2$O$_5$ consists of quantum spins similar to those in other low-dimensional quantum magnet [29], it is quintessential to explore a new universality class determined by exotic global properties.

In conclusion, we find a new quasi-one-dimensional chain system NiTe$_2$O$_5$, in which edge-shared NiO$_6$ octahedra form a chain and the chains are arranged in the distorted square-lattice. The temperature dependences of magnetic susceptibility and specific heat demonstrate an antiferromagnetic phase transition at ~ 30.5 K, exhibiting an archetypal anisotropy of magnetic susceptibility and lambda-shaped anomaly of specific heat. The single-crystal/powder neutron diffraction experiment reveals that the long-range antiferromagnetic order of NiTe$_2$O$_5$ has the interesting spin configuration of ferromagnetically ordered longitudinal magnetic moments, but alternating ferromagnetic-antiferromagnetically ordred transverse magnetic moments along the chain. Remarkably, the temperature dependence of single-crystal neutron diffraction clearly demonstrates that the antiferromagnetic order parameter develops with the unconventional critical exponent of $\alpha' \sim 0.25$ and $\beta \sim 0.18$ in the quasi-one-dimensional spin chain struture. Our experimental discovery presents rich unconventional critical behavior in a low-dimensional quantum magnet and provides a new facet in the research of quantum/topological magnetic systems.

**Acknowledgements**
Work at Ulsan National Institute of Science and Technology was supported by Basic Science Research Program through the National Research Foundation of Korea(NRF) funded by the Ministry of Science, ICT & Future Planning (NRF-2015R1C1A1A01055964, NRF-2016M2B2A4912417). Work at Florida carried out under the auspices of the U. S. Department of Energy, Basic Energy Sciences, contract no. DE-FG02-ER45268. This work at the IBS CCES and SNU was supported by the Institute for Basic Science (IBS) in Korea (IBS-R009-G1). The work at ORNL's HFIR was sponsored by the Scientific User Facilities Division, Office of Science, Basic Energy Sciences (BES), U.S. Department of Energy (DOE).

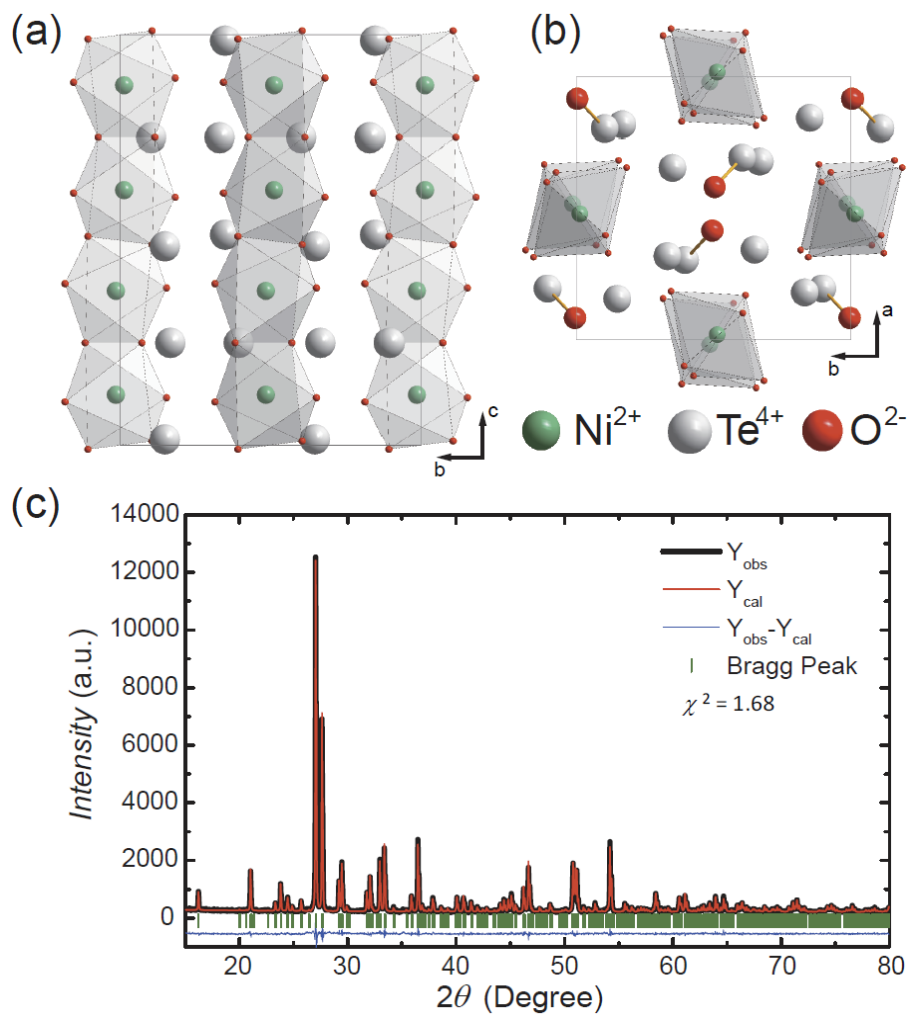

Figure 1. Crystallographic structure of $NiTe_2O_5$ in (a) *bc*-plane and (b) *ab*-plane. (c) Powder X-ray diffraction results (black line) at room temperature. The red line is the best fit from the Rietveld refinement using FULLPROF [20]. The green hash marks depict the position of Bragg peaks, and the bottom blue line represents the difference between the observed and calculated intensity.

| Chemical formula | NiTe$_2$O$_5$ |
|---|---|
| Structure | Orthorhombic |
| Space group | *Pbnm* |
| $a$ (Å) | 8.4418 |
| $b$ (Å) | 8.8663 |
| $c$ (Å) | 12.1203 |
| $R_P$ (%) | 14.2 |
| $R_{wp}$ (%) | 13.8 |
| $R_{exp}$ (%) | 10.6 |
| $\chi^2$ | 1.68 |
| Te1 ($x, y, z$) | (0.15962, 0.85151, 0.48628) |
| Te2 ($x, y, z$) | (0.19652, 0.10629, 0.25) |
| Te3 ($x, y, z$) | (0.68169, 0.33474, 0.25) |
| Ni ($x, y, z$) | (0.98462, 0.51679, 0.12273) |
| O1 ($x, y, z$) | (0.66773, 0.12569, 0.25) |
| O2 ($x, y, z$) | (0.12455, 0.45826, 0.25) |
| O3 ($x, y, z$) | (0.40877, 0.48597, 0.13537) |
| O4 ($x, y, z$) | (0.14059, 0.68765, 0.10950) |
| O5 ($x, y, z$) | (0.11526, 0.38428, 0.48515) |
| O6 ($x, y, z$) | (0.82248, 0.34407, 0.12771) |

**Table I. Unit cell parameters, reliability factors, and atomic positional parameters for NiTe$_2$O$_5$.**

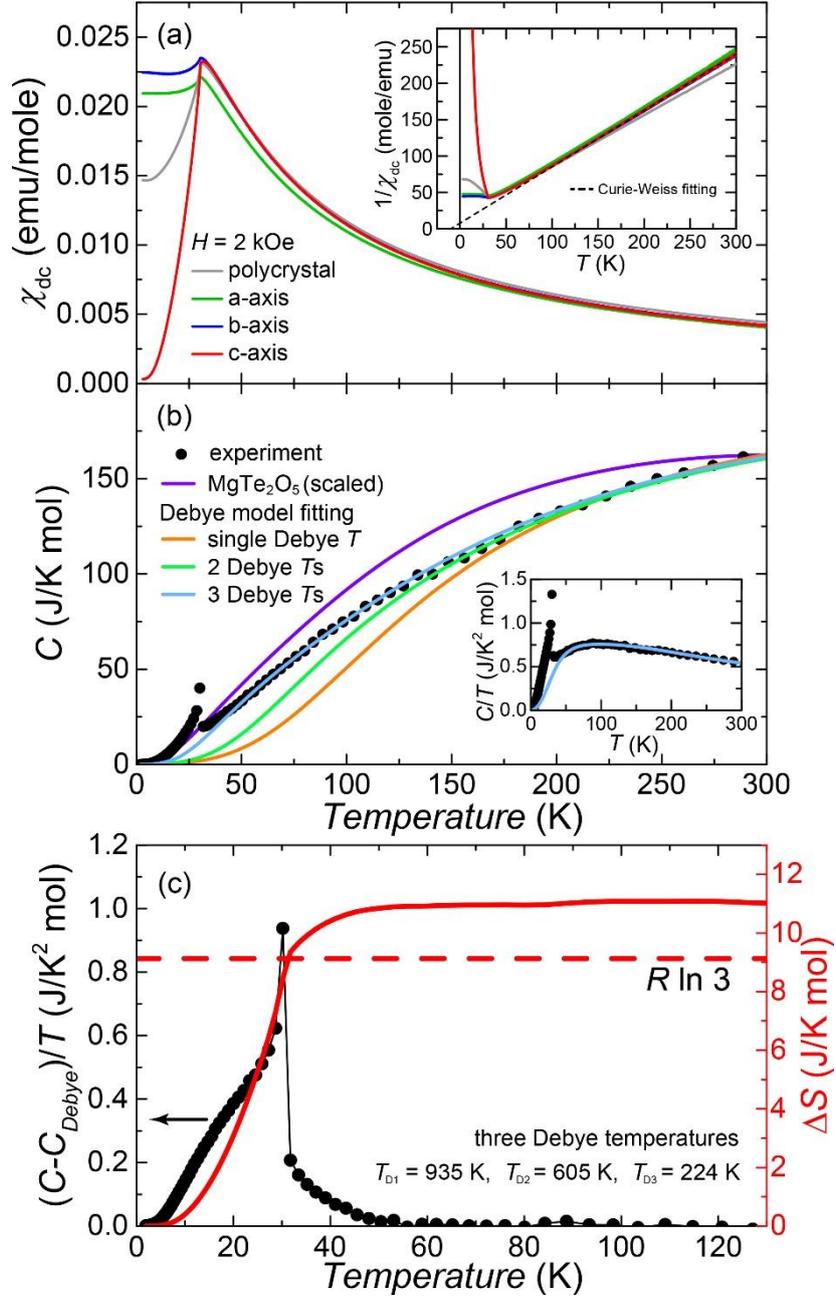

Figure 2. Temperature dependence of (a) the magnetic susceptibility of *a*- (green), *b*- (blue), *c*- (red) axis of single-crystalline and polycrystalline (black) NiTe$_2$O$_5$ in $H =$ 2 kOe, (b) the specific heat $C(T)$ of a single crystal. Orange, light green, and light blue curves indicate the fitting curve from the Debye model with single ($T_D = 613$ K), two ($T_{D1} = 1008$ K, $T_{D2} = 407$ K), and three ($T_{D1} = 935$ K, $T_{D2} = 605$ K, $T_{D3} = 224$ K) Debye temperatures, respectively. The purple curve represents scaled $C(T)$ of MgTe$_2$O$_5$ with the Lindemann factor [23,24]. Temperature dependence of inverse magnetic susceptibility and $C/T$ are shown in the insets of (a) and (b), respectively. A black dashed line indicates Curie-Weiss fitting result from 200 K to 350 K. (c) Temperature dependence of magnetic specific heat $C_{mag}/T$ (black solid circles) and magnetic entropy (red solid line). $C_{mag}$ is estimated by subtracting the phonon specific heat from the Debye model fitting of three Debye temperatures.

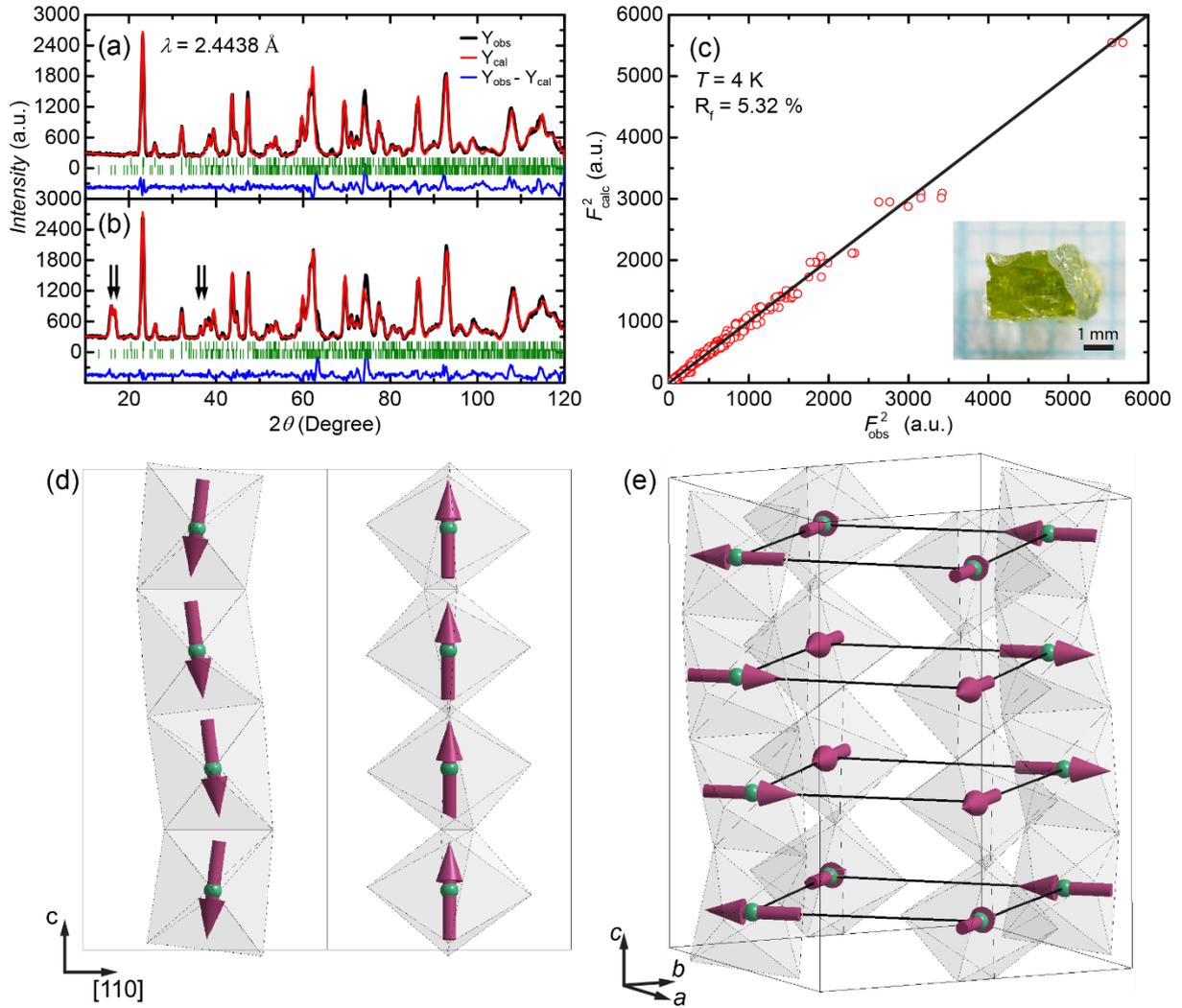

**Figure 3** Powder neutron diffraction patterns at (a) 300 K, and (b) 2 K. Black and red lines are the observed and calculated results, respectively. The upper and lower sets of hash marks signify the Bragg reflections of the nuclear and magnetic phase, respectively. In (b), the positions (from left to right) of the magnetic reflections (010), (100), (201), (120), (013), and (210) are marked by arrows. (c) Observed versus calculated intensity (square of the structure factor) of the magnetic Bragg reflections for NiTe$_2$O$_5$ single crystal in arbitrary units. Inset is a photo of the employed NiTe$_2$O$_5$ single crystal. Schematic magnetic structure of NiTe$_2$O$_5$ with (d) full magnetic moment of Ni$^{2+}$ in the c-[110] plane and (e) transverse component $m_\perp = \sqrt{m_a^2 + m_b^2}$ of the magnetic moment from the experimental result of single-crystal neutron diffraction.

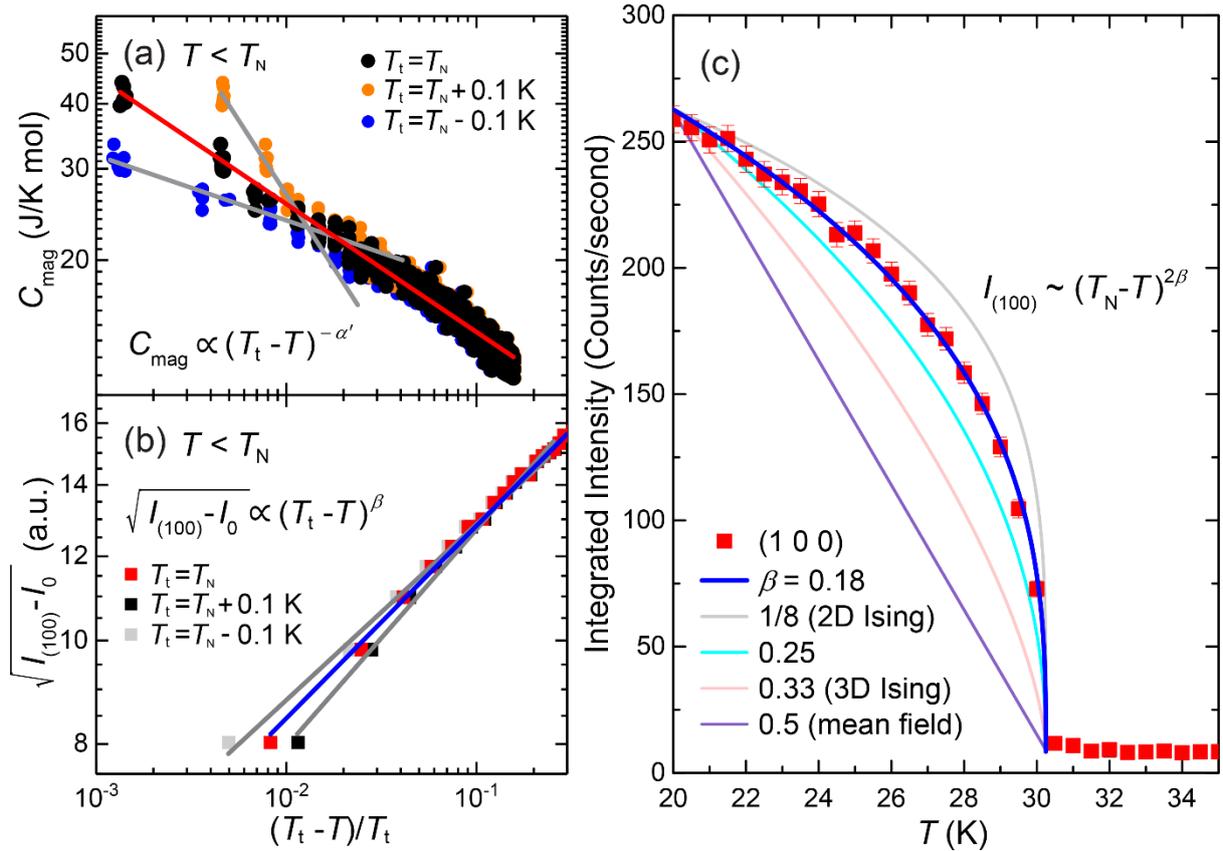

**Figure 4** Power-law fittings for the critical phenomena in (a) magnetic specific heat (red line) and (b) (100) magnetic reflection intensity $I_{(100)}$ (blue line) on logarithmic scales. Considering the experimental error of the transition temperature $T_t$, the power-laws are analyzed for $T_t = T_N − 0.1, T_N, T_N + 0.1$ K. The fittings give the critical exponents of $\alpha' = 0.14, 0.25, 0.31$ and $\beta = 0.17, 0.18, 0.20$ with respect to $T_t = T_N − 0.1, T_N, T_N + 0.1$ K, respectively. Background of (100) magnetic reflection $I_0 = 8.43733$ is estimated from a mean value of $I_{(100)}$ at $T_N < T < 35$ K. (c) Temperature evolution of the (100) magnetic reflection intensity below $T_N$. The blue solid line is the best fit to the power law, $I_{(100)} \sim (T_N − T)^{2\beta}$, where $\beta = 0.18$ is the critical exponent. Gray, cyan, pink, and purple solid lines are power law behavior for $\beta = 1/8$ (2D Ising), $0.25$, $0.33$ (3D Ising), and $0.5$ (mean field), respectively.

# References

[1] L. P. Kadanoff, W. Götze, D. Hamblen, R. Hecht, E. A. S. Lewis, V. V. Palciauskas, M. Rayl, J. Swift, D. Aspnes, and J. Kane, Rev. Mod. Phys. **39**, 395 (1967).
[2] O. Peters, and J. D. Neelin, Nat. Phys. **2**, 393 (2006).
[3] I. Hetel, T. R. Lemberger, and M. Randeria, Nat. Phys. **3**, 700 (2007).
[4] L. Zhang, and F. Wang, Phys. Rev. Lett. **118**, 087201 (2017).
[5] P. Merchant, B. Normand, K. W. Krämer, M. Boehm, D. F. McMorrow, Ch. Rüegg, Nat. Phys. **10**, 373 (2014).
[6] A. Narayan, A. Cano, A. V. Balatsky, N. A. Spaldin, Nat. Mater. **18**, 223 (2019).
[7] S. V. Kravchenko, W. E. Mason, G. E. Bowker, J. E. Furneaux, V. M. Pudalov, M. D'Iorio, Phys. Rev. B **51**, 7038 (1995).
[8] P. V. Lin, and D. Popovic, Phys. Rev. Lett. **114**, 166401 (2015).
[9] H. E. Stanley, Rev. Mod. Phys. **71**, S358 (1999).
[10] W. Knafo, C. Meingast, S. Sakarya, N. H. van Dijk, Y. Huang, H. Rakoto, J.-M. Broto, H. v. Löhneysen, J. Phys. Soc. Jpn. **78**, 043707 (2009).
[11] N. Tateiwa, Y. Haga, T. D. Matsuda, E. Yamamoto, Z. Fisk, Phys. Rev. B **89**, 064420 (2014).
[12] T. Kida, A. Senda, S. Yoshii, M. Hagiwara, T. Takeuchi, T. Nakano, I. Terasaki, EPL (Europhysics Letters) **84**, 27004 (2008).
[13] A. Omerzu, M. Tokumoto, B. Tadic, D. Mihailovic, Phys. Rev. Lett. **87**, 177205 (2001).
[14] D. Fuchs, M. Wissinger, J. Schmalian, C. L. Huang, R. Fromknecht, R. Schneider, H. v Löhneysen, Phys. Rev. B **89**, 174405 (2014).
[15] N. D. Mermin, and H. Wagner, Phys. Rev. Lett. **17**, 1133 (1966).
[16] K. Wierschem, and P. Sengupta, Phys. Rev. Lett. **112**, 247203 (2014).
[17] A. C. Walters, T. G. Perring, J.-S. Caux, A. T. Savici, G. D. Gu, C.-C. Lee, W. Ku, I. A. Zaliznyak, Nat. Phys. **5**, 867 (2009).
[18] G. Xu, J. F. DiTusa, T. Ito, K. Oka, H. Takagi, C. Broholm, G. Aeppli, Phys. Rev. B **54**, R6827 (1996).
[19] C. Platte, and M. Trömel, Acta Crystallogr. Sect. B-Struct. Sci. Cryst. Eng. Mat. **37**, 1276 (1981).
[20] J. Rodríguez-Carvajal, Phys. B **192**, 55 (1993).
[21] See Supplemental Material at "URL" for detailed description of the data analysis of phonon contribution in specific heat
[22] M. Troemel, Zeitschrift fuer Anorganische und Allgemeine Chemie **418**, 141 (1975).
[23] M. Bouvier, P. Lethuillier, and D. Schmitt, Phys. Rev. B **43**, 13137 (1991).
[24] F. Lindemann, Z. Phys. **11**, 609 (1910).
[25] J. M. Perez-Mato, S. V. Gallego, E. S. Tasci, L. Elcoro, G. de la Flor, and M.I. Aroyo, Annu. Rev. Mater. Res. **45**, 217 (2015).
[26] A. Taroni, S. T. Bramwell, and P. C. Holdsworth, J. Phys. Condens. Matter **20**, 275233 (2008).
[27] D. Kim, B. Revaz, B. L. Zink, F. Hellman, J. J. Rhyne, J. F. Mitchell, Phys. Rev. Lett. **89**, 227202 (2002).
[28] F. Kagawa, K. Miyagawa, and K. Kanoda, Nature **436**, 534 (2005).
[29] A. Vasiliev, O. Volkova, E. Zvereva, M. Markina, npj Quantum Materials **3**, 18 (2018).